\documentclass[aps,floatfix,showpacs]{revtex4}
\usepackage{graphics,epsfig}

\begin{document}

\title{Invariants of the Kerr Vacuum}
\author{Kayll Lake \cite{email}}
\affiliation{Department of Physics and Department of Mathematics
and Statistics, Queen's University, Kingston, Ontario, Canada, K7L
3N6 }
\date{\today}

\begin{abstract}
The Kerr vacuum has two independent invariants derivable from the
Riemann tensor without differentiation.  Both of these invariants
must be examined in order to avoid an erroneous conclusion that
the ring singularity of this spacetime is ``directional".
\end{abstract}


\maketitle Recently Schmidt \cite{schmidt} has reminded us, by way
of a simple example, that the square of the Weyl tensor can be
negative. This behavior is a well known property of the Kerr
vacuum \cite{brad}, \cite{new}, \cite{henry} in which the Weyl
tensor becomes highly anisotropic sufficiently close to the
singularity. Indeed, the divergence of this scalar (equivalent
here to the Kretschmann scalar) is a standard way in which texts
point out the singularity in this spacetime (e.g.
\cite{hawkingellis}, \cite{wald}, \cite{dinverno}) though an
explicit form of this scalar is a little hard to find in the texts
(correct forms are give in \cite{defeliceclarke} and
\cite{poisson} and an incorrect form is given in
\cite{ludvigsen}). The scalar is not difficult to calculate by
hand (using, for example, the Newman-Penrose formalism) and so it
not a surprise that it was trivial to calculate (in a radiating
Kerr-Newman generalization) on a personal computer a decade ago
\cite{musgrave}. The Kerr vacuum has two independent invariants
derivable from the Riemann tensor without differentiation. In this
note we examine each in detail and point out that both must be
examined in order to avoid an erroneous conclusion that the ring
singularity is in any sense ``directional".

\begin{figure}
\epsfig{file=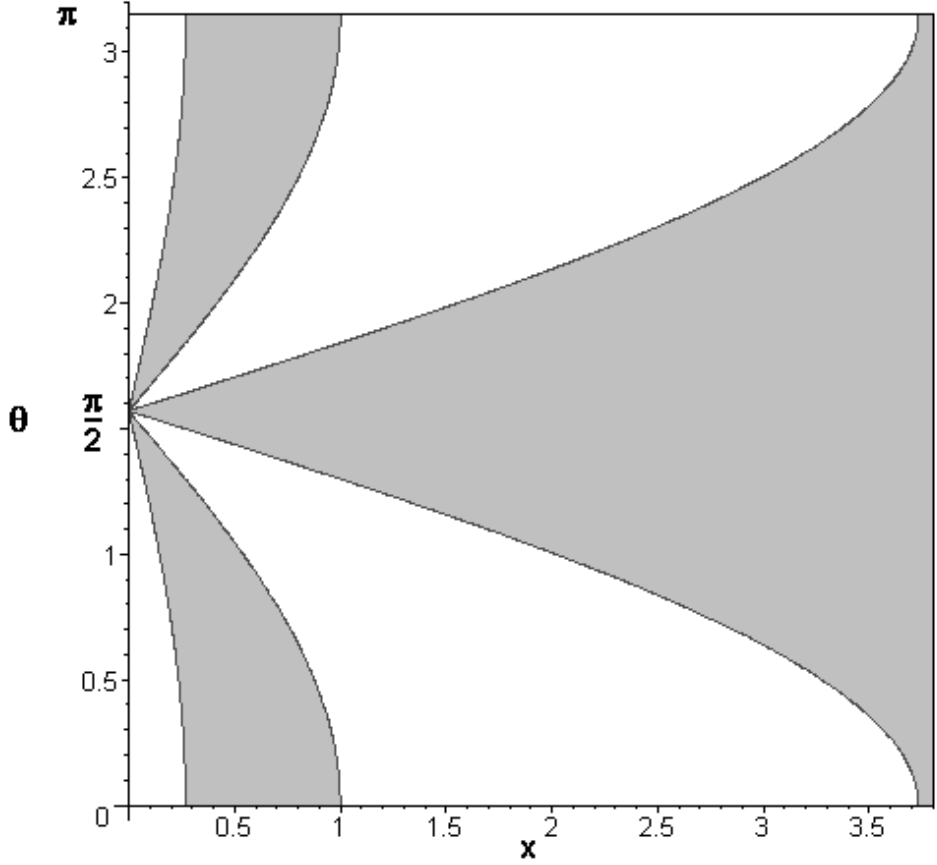,height=5in,angle=0}
\caption{\label{signw}Regions with $\mathcal{W}(\equiv \frac{C_{i
j k l}C^{i j k l}a^6}{48m^2})>0$ are shaded. The shaded region
extends indefinitely to the right where $\mathcal{W} \rightarrow
0$ as $x \rightarrow \infty$. The boundaries $\mathcal{W}=0$ are
given by $x = \pm x1,\;  \pm x2,\; \pm x3$ as explained in the
text. (In the corresponding Kerr-Schild representation the
boundaries reduce to circles, two of radius 2 centered on $\pm
\sqrt{3}$ and one of radius 1 centered at the origin. For our
purposes here we find the $\theta - x$ plane more useful.) }
\end{figure}

\begin{figure}
\epsfig{file=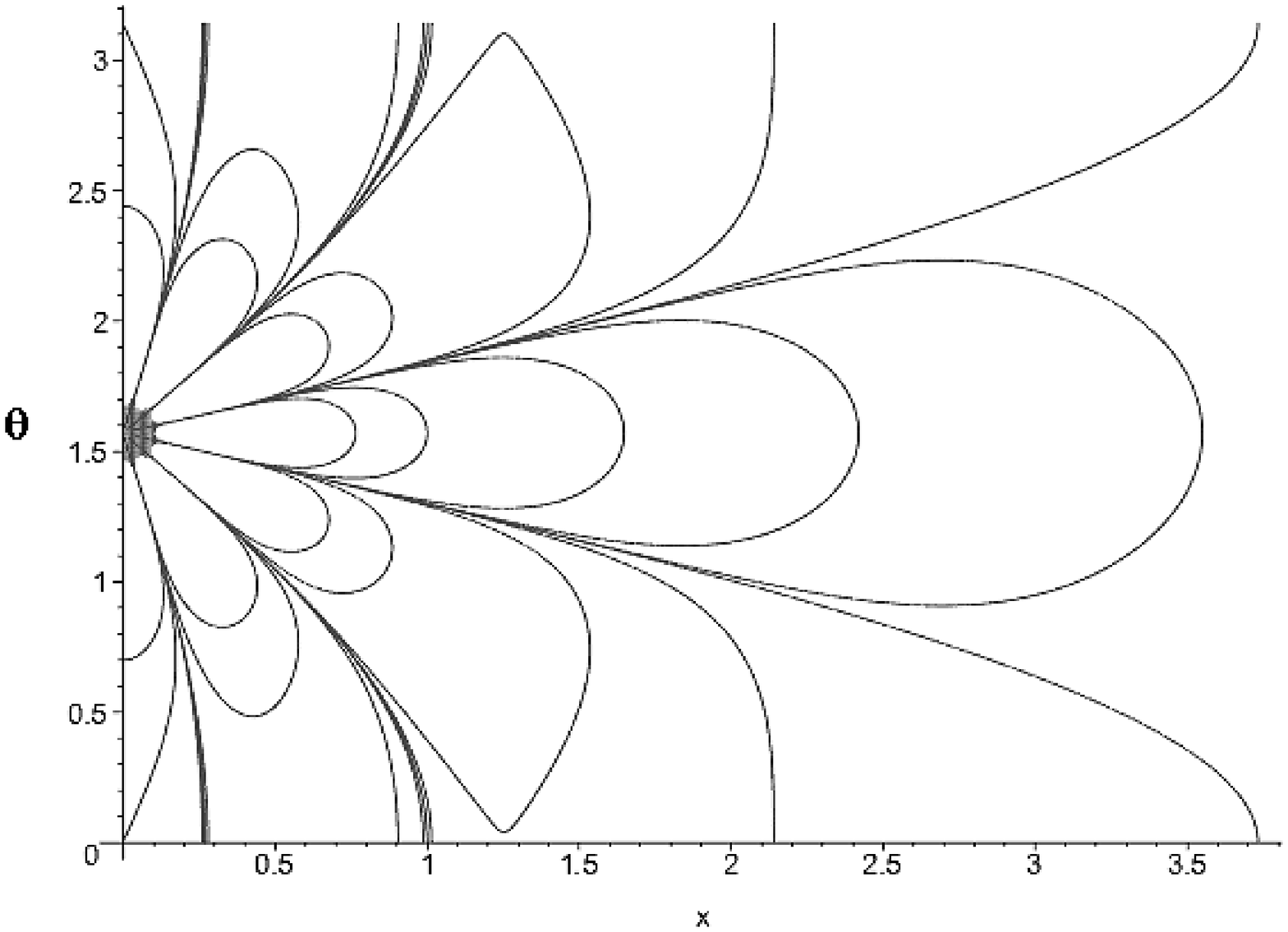,height=4.5in,angle=0}
\caption{\label{consw} Level curves of constant $\mathcal{W}$. The
values of $\mathcal{W}$ shown are 5, 1, 0.05, 0.005, 0.0005, 0,
-0.005, -0.0368, -1 and -5.}
\end{figure}

\begin{figure}
\epsfig{file=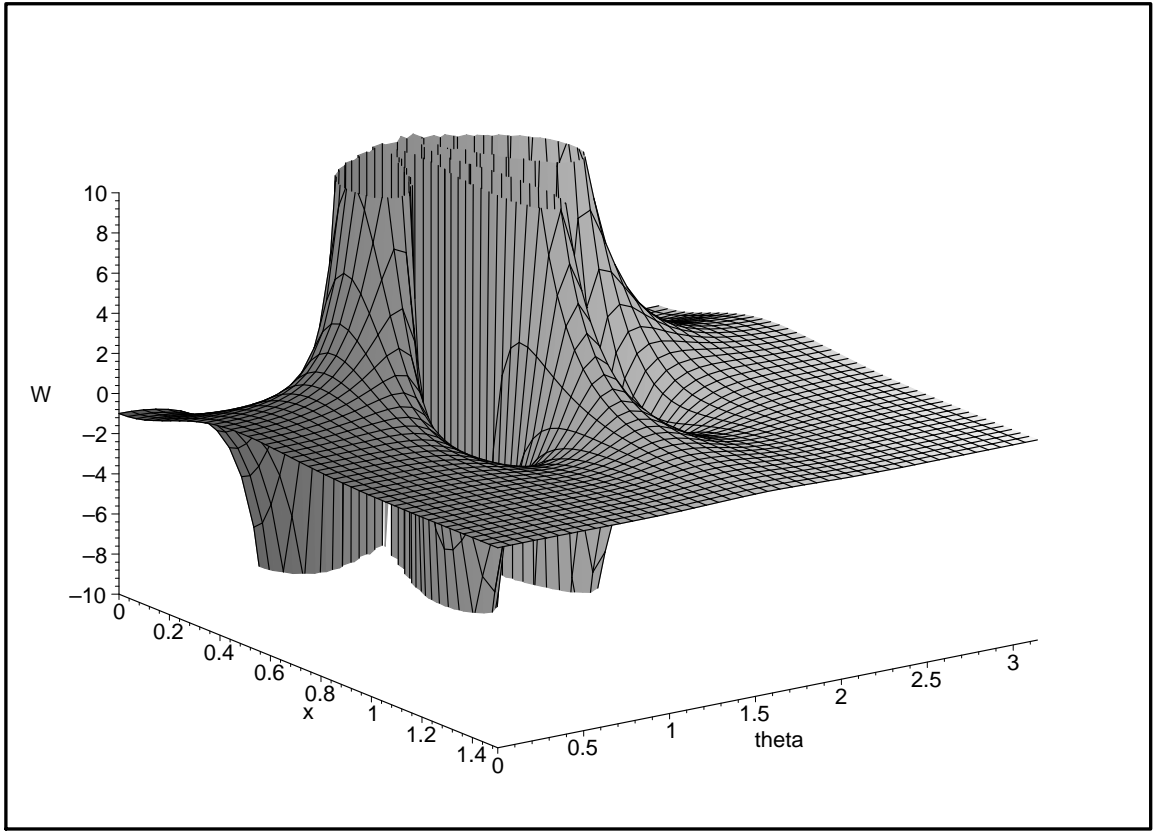,height=6in,angle=-90}
\caption{\label{wtop}Plot of $\mathcal{W}(\equiv \frac{C_{i j k
l}C^{i j k l}a^6}{48m^2})$ showing a variation in both
$x\equiv\frac{r}{a}$ and $\theta$.}
\end{figure}

\begin{figure}
\epsfig{file=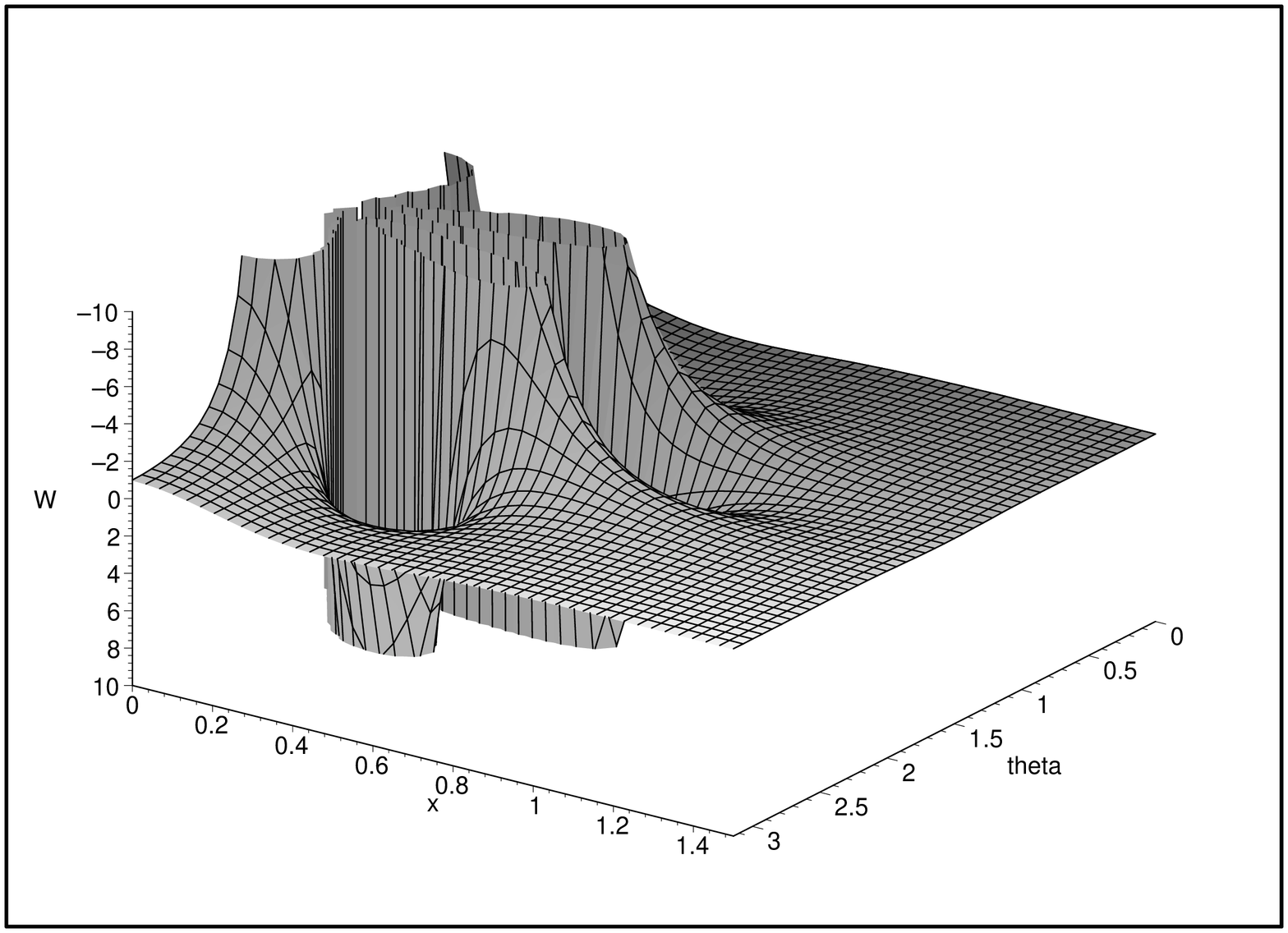,height=6in,angle=-90}
\caption{\label{wbottom}Plot of $\mathcal{W}(\equiv \frac{C_{i j k
l}C^{i j k l}a^6}{48m^2})$ showing a variation in both
$x\equiv\frac{r}{a}$ and $\theta$. This is the underside of Figure
\ref{wtop}. }
\end{figure}

\bigskip
In terms of the familiar Boyer-Lindquist coordinates \cite{boyer},
writing $x \equiv r/a, a \neq 0$ and $W \equiv C_{i j k l} C^{i j
k l}$ where $C_{i j k l}$ is the Weyl tensor, it follows that
\begin{equation}
\mathcal{W}\equiv \frac{Wa^6}{48m^2}={\frac { ( x- x1 ) ( x+ x1) (
x-x2) ( x+x2) ( x-x3 )  ( x+x3) }{ ( x^{2}+x1^{2} ) ^{6}}}
\label{weylsq}
\end{equation}
where $x1 \equiv cos(\theta),\; x2 \equiv
(2+\sqrt{3})\;cos(\theta)$ and $x3 \equiv
(2-\sqrt{3})\;cos(\theta)$. Note that (\ref{weylsq}) holds for all
$a \neq 0$ and $m \neq 0$. Along $x=0$,
$\mathcal{W}\;cos(\theta)^6=-1$ but along $\theta=\pi/2,
\mathcal{W}\;x^6=+1$ and so $\mathcal{W}$ fails to be continuous
in the limit $x \rightarrow 0,\; \theta \rightarrow \pi/2$
($\equiv \mathcal{S}$). Indeed $\mathcal{W}$ need not diverge in
the limit $\mathcal{S}$ along trajectories asymptotic to $x = \pm
x1,  \pm x2, \pm x3$. The sign of $\mathcal{W}$ is summarized in
Figure \ref{signw} and some level curves of constant $\mathcal{W}$
are shown in Figure \ref{consw}, both in the $\theta-x$ plane. The
function $\mathcal{W}$ is plotted in Figures \ref{wtop} and
\ref{wbottom}. This extends the analysis in \cite{brad} and in
\cite{henry} (the special case given in Figure 2 of \cite{henry}
is incomplete).

\begin{figure}
\epsfig{file=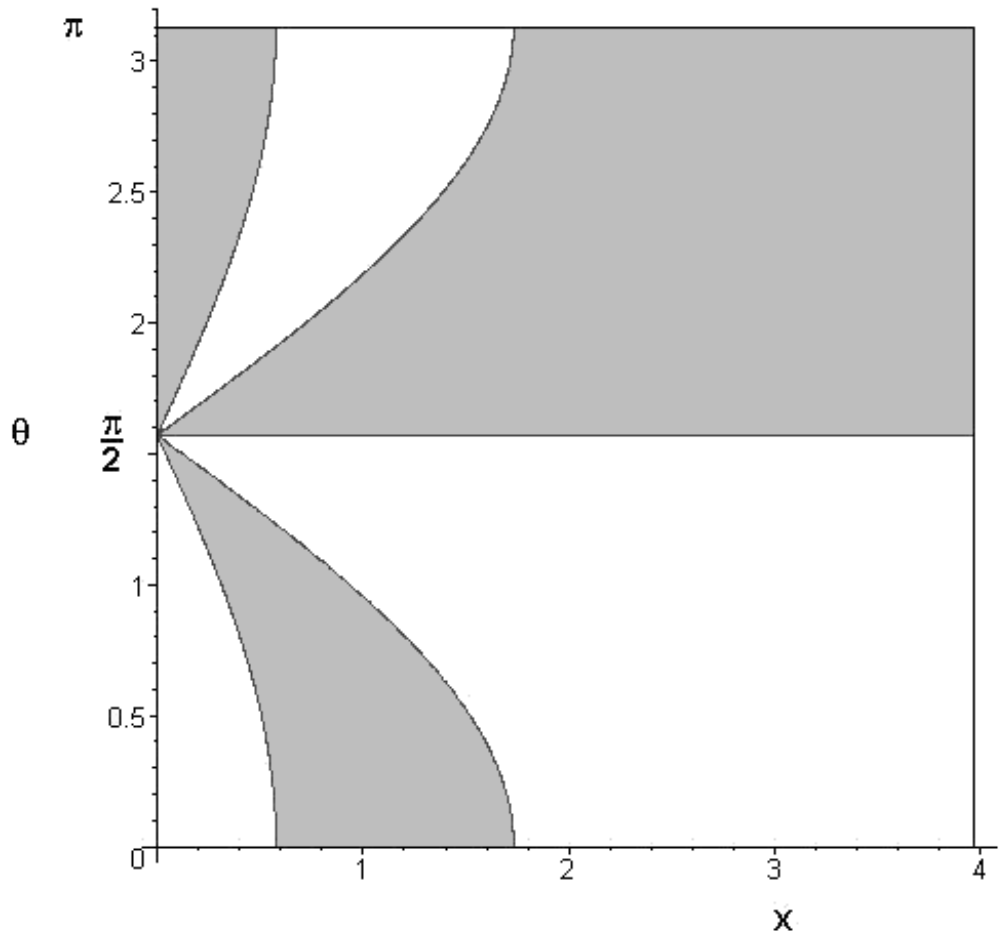,height=5in,angle=0}
\caption{\label{signwstar}Regions with ${\mathcal{^{*}W}}(\equiv
\frac{^{*}C_{i j k l}C^{i j k l}a^6}{96m^2})>0$ are shaded. The
boundaries $\mathcal{^{*}W}=0$ are given by $x=0$, $\theta=\pi/2$,
$x = \pm x4$ and $x= \pm x5$ as explained in the text.}
\end{figure}

\begin{figure}
\epsfig{file=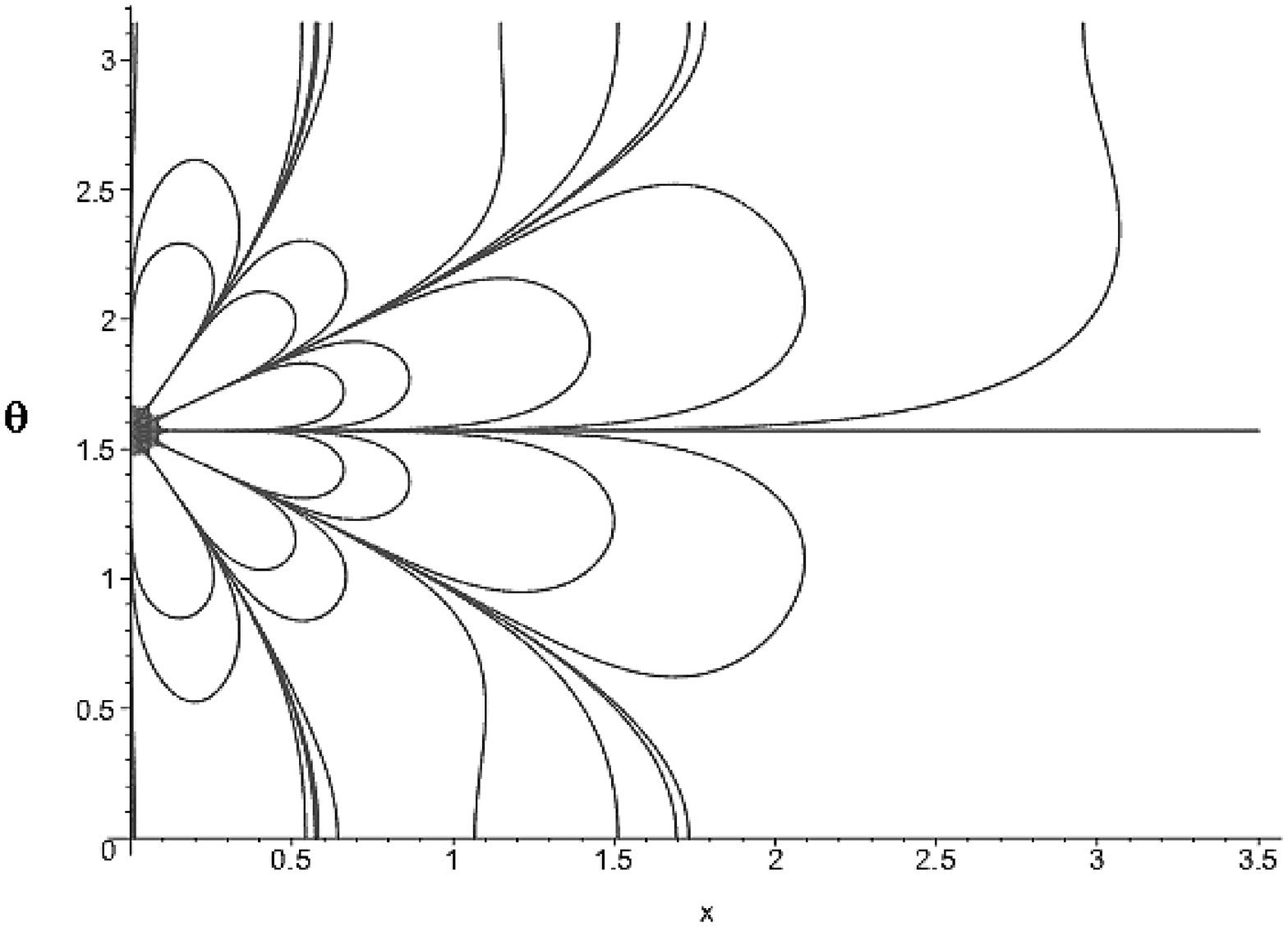,height=4.5in,angle=0}
\caption{\label{conswstar}Level curves of constant
$\mathcal{^{*}W}$. The values of $\mathcal{^{*}W}$ shown are 5, 1,
0.05, 0.005, 0.0005, 0, -0.005, -0.0368, -1 and -5.}
\end{figure}

\begin{figure}
\epsfig{file=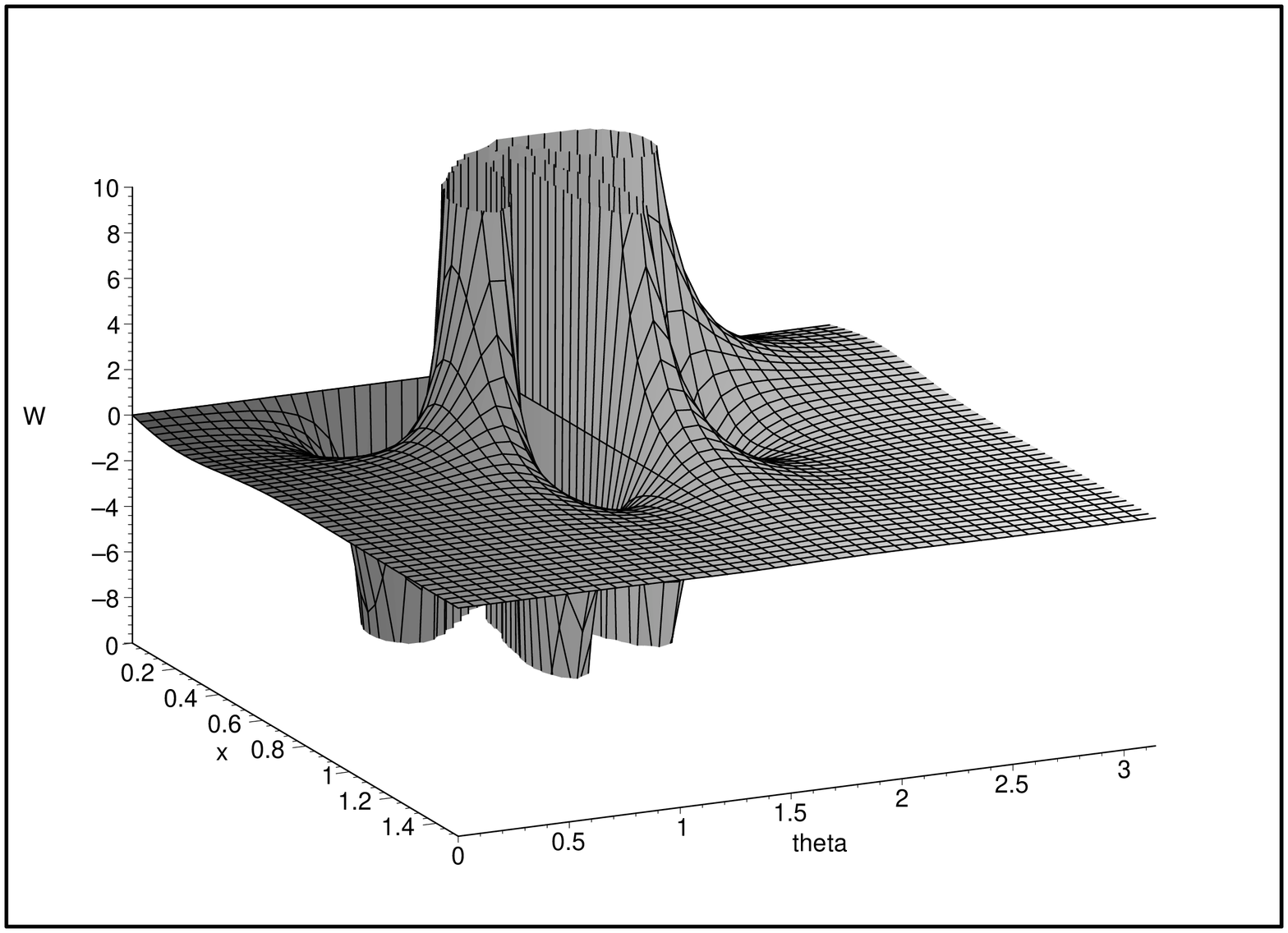,height=6in,angle=-90}
\caption{\label{wstartop}Plot of ${\mathcal{^{*}W}}(\equiv
\frac{^{*}C_{i j k l}C^{i j k l}a^6}{96m^2})$ showing a variation
in both $x\equiv\frac{r}{a}$ and $\theta$.}
\end{figure}

\begin{figure}
\epsfig{file=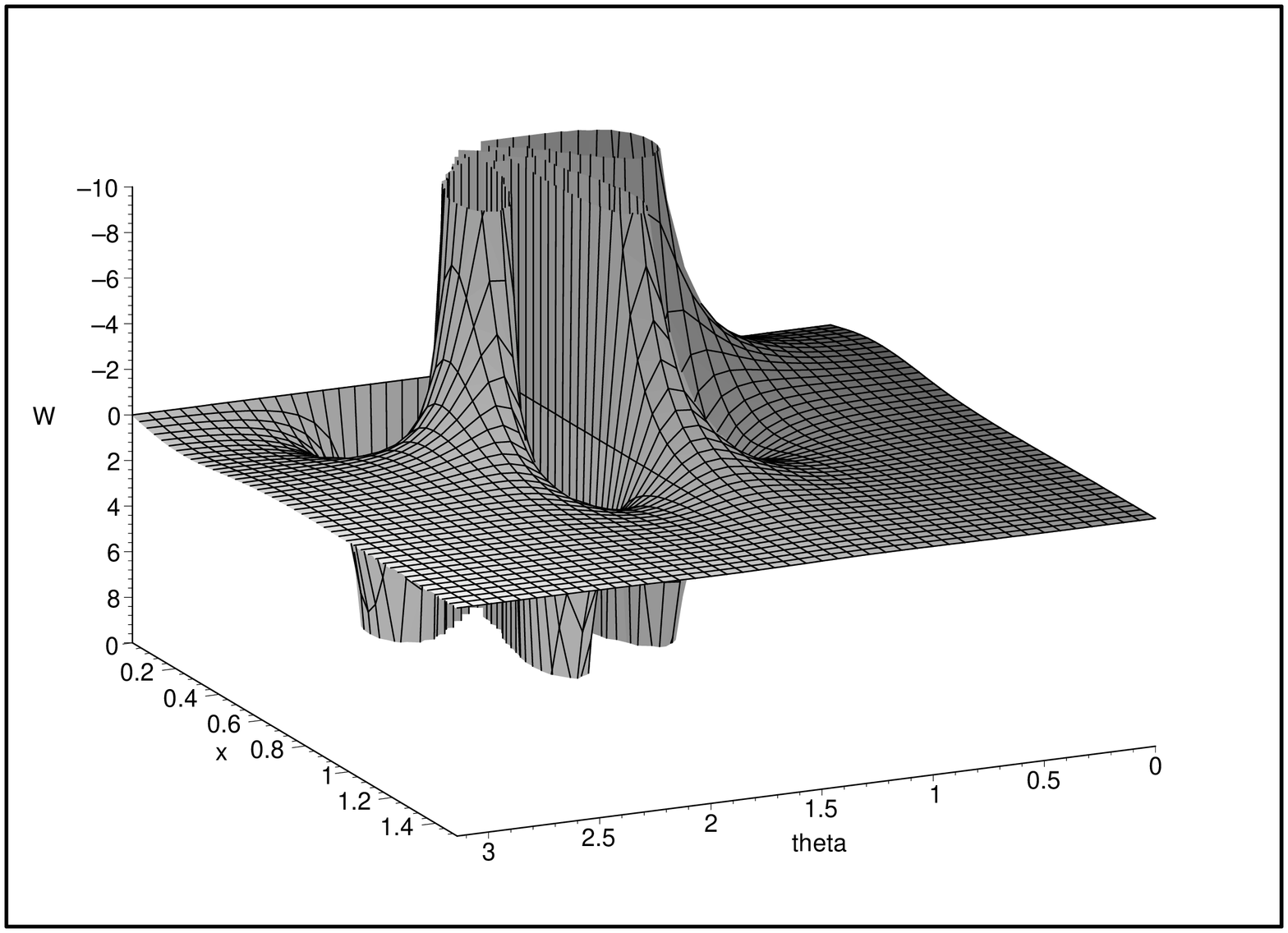,height=6in,angle=-90}
\caption{\label{wstarbottom}Plot of ${\mathcal{W}}(\equiv
\frac{^{*}C_{i j k l}C^{i j k l}a^6}{96m^2})$ showing a variation
in both $x\equiv\frac{r}{a}$ and $\theta$. This is the underside
of Figure \ref{wstartop}. }
\end{figure}

\bigskip
Any vacuum solution of Einstein's equations has only one further
independent quadratic invariant, $^{*}W \equiv \;^{*}C_{i j k l}
C^{i j k l}$ where $^{*}C_{i j k l}$ is dual to the Weyl tensor.
(See, for example, \cite{weinberg}. $W$ and $^{*}W$ are equivalent
to the real and imaginary parts of the complex Weyl scalar in
spinor notation. In \cite{new} $\;^{*}W$ is called the
Chern-Pontryagin invariant.) With $x$ and $x1$ defined as in
(\ref{weylsq}) it follows that
\begin{equation}
\mathcal{^{*}W}\equiv \frac{^{*}Wa^6}{96m^2}={\frac { -3x x1(
x-x4) ( x+x4) ( x-x5 )  ( x+x5) }{ ( x^{2}+x1^{2} ) ^{6}}}
\label{weylstarsq}
\end{equation}
where $x4 \equiv \sqrt{3}\;cos(\theta)$ and $x5 \equiv
cos(\theta)/\sqrt{3}$. There are no further independent
(non-differential) invariants in the Kerr vacuum (as it is of
Petrov type D). Again note that (\ref{weylstarsq}) holds for all
$a \neq 0$ and $m \neq 0$. Now $\mathcal{^{*}W}=0$ along $x=0$
but, for example, $16 \mathcal{^{*}W}\;cos(\theta)^6=+1$ along
$x=x1$ and so $\mathcal{^{*}W}$ also fails to be continuous in the
limit $\mathcal{S}$. $\mathcal{^{*}W}=0$ along $x=0$ and
$\theta=\pi/2$ and need not diverge in the limit $\mathcal{S}$
along trajectories asymptotic to $x = \pm x4$ and $\pm x5$. The
sign of $\mathcal{^{*}W}$ is summarized in Figure \ref{signwstar}
and some level curves of constant $\mathcal{^{*}W}$ are shown in
Figure \ref{conswstar}, again both in the $\theta-x$ plane. The
function $\mathcal{^{*}W}$ is plotted in Figures \ref{wstartop}
and \ref{wstarbottom}.

\bigskip
Whereas there exist trajectories (albeit non-geodesic) along which
$\mathcal{W}$ or $\mathcal{^{*}W}$ do not diverge in the limit
$\mathcal{S}$, there exists no trajectory along which both
$\mathcal{W}$ and $\mathcal{^{*}W}$ remain finite in this limit.
In this sense the singularity of the Kerr vacuum is not
directional.

\bigskip
Dynamic images of $\mathcal{W}$ and $\mathcal{^{*}W}$ are
available at \texttt{http://grtensor.org/negweyl} the details of
which are difficult to summarize in static images.
\begin{acknowledgments}
This work was supported by a grant from the Natural Sciences and
Engineering Research Council of Canada. Portions of this work were
made possible by use of \textit{GRTensorII} \cite{grt}.
\end{acknowledgments}

\newpage

\end{document}